%% file: main.tex
\documentclass[conference, 10pt]{IEEEtran}

\pdfoutput=1

\usepackage{blindtext, graphicx}
\usepackage[table,xcdraw]{xcolor}

\usepackage{listings}
\usepackage{epigraph}
\usepackage{float}
\usepackage{subfig}
\lstdefinelanguage{scala}{
    morekeywords={abstract,case,catch,class,def,%
        do,else,extends,false,final,finally,%
        for,if,implicit,import,match,mixin,%
        new,null,object,override,package,%
        private,protected,requires,return,sealed,%
        super,this,throw,trait,true,try,%
        type,val,var,while,with,yield},
    otherkeywords={=>,<-,<\%,<:,>:,\#,@},
    sensitive=true,
    morecomment=[l]{//},
    morecomment=[n]{/*}{*/},
    morestring=[b]",
    morestring=[b]',
    morestring=[b]"""
}

\newif\ifusebiblatex

\usebiblatexfalse

\ifusebiblatex
\usepackage[style=ieee, natbib=true]{biblatex}
\fi

\newif\ifshowcomments

\showcommentsfalse

\ifshowcomments
\newcommand{\mynote}[2]{\fbox{\bfseries\sffamily\scriptsize{#1}}
 {\small$\blacktriangleright$\textsf{\emph{#2}}$\blacktriangleleft$}}
\else
\newcommand{\mynote}[2]{}
\fi

\newcommand{\conor}[1]{\mynote{Conor}{#1}}
\newcommand{\svetozar}[1]{\mynote{Svetozar}{#1}}

\ifCLASSINFOpdf
\else
\fi
\begin{document}
%
\title{DINAMITE: A modern approach to memory performance profiling}

\newif\ifshowauthors

\showauthorstrue

\ifshowauthors
\author{\IEEEauthorblockN{Svetozar Miucin}
\IEEEauthorblockA{Electrical and Computer Engineering\\
University of British Columbia\\
Vancouver, Canada\\
smiucin@ece.ubc.ca}
\and
\IEEEauthorblockN{Conor Brady}
\IEEEauthorblockA{School of Computing Science\\
Simon Fraser University\\
Burnaby, Canada\\
cbrady@sfu.ca}
\and
\IEEEauthorblockN{Alexandra Fedorova}
\IEEEauthorblockA{Electrical and Computer Engineering\\
University of British Columbia\\
Vancouver, Canada\\
sasha@ece.ubc.ca}}
\fi


%


\maketitle
\input{abstract}



%
\IEEEpeerreviewmaketitle


\lstdefinestyle{style1}{
  xleftmargin=2\parindent,
  belowcaptionskip=1\baselineskip,
  breaklines=true,
  frame=L,
  linewidth=0.9\linewidth,
  numbers=left,
  showstringspaces=false,
  basicstyle=\footnotesize\ttfamily
}
\parskip 0pt

\input{introduction}
\input{systemdesign}

\input{evaluation}
\input{relatedwork}

\input{futurework}
\input{machines}
\ifCLASSOPTIONcaptionsoff
  \newpage
\fi



%

%

\ifusebiblatex
\printbibliography
\else 
\bibliographystyle{IEEEtran}

\bibliography{IEEEabrv,bibliography}
\fi






%

\begin{IEEEbiography}[{\includegraphics[width=1in,height=1.25in,clip,keepaspectratio]{picture}}]{John Doe}
\blindtext
\end{IEEEbiography}




\end{document}

%% file: abstract.tex
\begin{abstract}

Diagnosing and fixing performance problems on multicore machines with deep memory hierarchies is extremely challenging. Certain problems are best addressed when we can analyze the entire trace of program execution, e.g., every memory access. Unfortunately such detailed execution logs are very large and cannot be analyzed by direct inspection. We present DINAMITE: a toolkit for \textbf{D}ynamic \textbf{IN}strumentation and \textbf{A}nalysis for \textbf{M}ass\textbf{I}ve \textbf{T}race \textbf{E}xploration. DINAMITE is a collection of tools for end-to-end performance analysis: from the LLVM compiler pass that instruments the program to plug-and-play tools that use a modern data analytics engine Spark Streaming for trace introspection.  Using DINAMITE we found opportunities to improve data layout in several applications that resulted in 15-20\% performance improvements and found a shared-variable bottleneck in a popular key-value store, whose elimination improved performance by 20x.

\end{abstract}

%% file: introduction.tex
\section{Introduction}
 
Memory performance is a limiting factor in many important programs. Traditional database systems, web servers, scientific algorithms and modern data analytics programs alike were observed to spend 50-80\% of CPU cycles stalled on memory \cite{ailamaki1999dbmss} \cite{ferdman2012clearing}. That is, 50-80\% of the time these programs are unable to commit any instructions due to outstanding long-latency memory accesses.  Understanding and addressing the causes of these bottlenecks is of paramount importance. Performance improvements from a more efficient memory layout or improved locality of access are usually significant and in some cases reach an order of magnitude \cite{lattner2005automatic} \cite{lachaize2012memprof} \cite{pesterev2010locating} \cite{boyd2010analysis} \cite{goto2008anatomy} \cite{yoon2005cache}.

At the same time, optimizing memory performance is notoriously difficult. Compiler optimizations can be effective when static analysis is sufficient to infer improvement opportunities. However, the scope of static optimizations is limited \cite{bruening2003infrastructure}, partly because insight into a bottleneck can often be gained only during execution and partly because the compiler is limited in how it can change data structure layout, particularly with dynamically allocated data structures and in unmanaged languages. As a result, developers often resort to manually optimizing their data structures and algorithms, relying on tools for dynamic program analysis and memory profiling.

Unfortunately, most existing tools suffer from either lack of generality, portability, or flexibility. Conventional CPU profilers, such as perf \cite{de2010new}, aim to identify source locations that generate the majority of cache misses, but because of skid effects in hardware counters \cite{bitzes2014overhead}, \cite{itzkowitz2003memory} or compiler optimizations, such as function inlining, this information is often imprecise or plain wrong. Cachegrind \cite{nethercote2007valgrind} accurately identifies source lines generating cache misses, but does not provide actionable insight that might lead the programmer to reduce them. Dprof \cite{pesterev2010locating} identifies data structures and fields that are responsible for cache misses due to sharing among threads, but it is not flexible enough to address other causes of poor memory performance and was designed specifically for the Linux kernel. Similarly, Memprof \cite{lachaize2012memprof} focuses on identifying objects that cause remote accesses on NUMA systems, but the implementation is Linux-specific, tied to AMD hardware and does not lend itself to other types of analyses.

To address this gap, we built DINAMITE -- a toolkit for \textbf{D}ynamic \textbf{IN}strumentation and \textbf{A}nalysis for \textbf{M}ass\textbf{I}ve \textbf{T}race \textbf{E}xploration. DINAMITE uses compile-time instrumentation to inject tracing code into the program. At runtime, the program generates precise traces containing every memory access, its source location and the corresponding variable name, type and value. These traces are then used to perform various memory-related analyses, for example, identifying highest cache-miss offenders, locating hot and cold fields of a data structure, correlating locality of accesses with values of variables, detecting true and false sharing, building arbitrary models of memory access patterns, and many others. 

The approach of using instrumentation and tracing is by itself not new; it is used in Pin \cite{luk2005pin}, Valgrind/Cachegrind \cite{nethercote2007valgrind} and other similar tools. Its main downside is high runtime overhead and very large execution traces, which can reach hundreds of gigabytes even for small programs. However, for the very challenging task of memory performance debugging this approach is often the only practical option, because certain analyses, e.g., those relying on cache simulation, can be performed only with a precise execution trace.  

Key contributions of DINAMITE are as follows:
\vspace{-10pt}
\begin{itemize}
    \item The instrumentation is implemented as a pass in LLVM \cite{lattner2004llvm}, so it is applicable to any language with an LLVM front-end.
    \item Since the instrumentation is done at compile-time, the source-level debug information assigned to trace entries is precise and easy to extract.
    \item Traces are generated in binary format and with buffering, which is the most efficient method known to us. As a result, DINAMITE`s runtime overhead is similar or smaller than state-of-the-art instrumentation tools, like Pin and Valgrind. 
    \item DINAMITE gives the user flexibility in how to handle execution traces. The traces can be stored in the file system, but if the user does not wish or cannot store these typically large traces, they can be analyzed on-the-fly using a streaming analytics engine like Spark Streaming \cite{zaharia2012discretized} (or any other similar engine). \item DINAMITE is easy to extend with additional analysis tools. A developer can write a new tool with a few lines of Scala (if using DINAMITE with Spark) or any other language of choice. We target advanced developers who understand how software interacts with memory hierarchies of modern processors,  so we wanted to give them ultimate flexibility in analysing memory traces. 
\end{itemize}

We built three tools on top of DINAMITE. The first one produces variable names and source lines responsible for the highest number of cache accesses and misses. Using it, we reduced the last-level cache (LLC) miss rate of \textit{429.mcf} from SPEC2006 by 55\% and improved its performance by 12\%. We also reduced the LLC missrate and improved performance of PARSEC's \textit{fluidanimate} by 50\% and 15\% respectively. 

The second tool implements Chilimbi's structure splitting algorithm~\cite{chilimbi1999cache}. Thanks to it, we reduced the LLC miss rate of SPEC2006's \textit{429.mcf} by 60\%, with the corresponding 20\% reduction in runtime. 

The third tool detects program variables that are heavily shared by many threads. This tool enabled us to detect a previously known performance bottleneck in WiredTiger, MongoDB's back-end key/value store \cite{wt} \cite{wt-mongo-db} \cite{wt-pull-request}. Even though the bottleneck was already known and fixed before we created DINAMITE (in fact, this was one of the motivating reasons for DINAMITE), the original discovery took several weeks, while DINAMITE pin-pointed it in a few hours. Performance improvement of the read-only sequential LevelDB benchmark implemented over WiredTiger reached a factor of 20 for 32 threads.

The rest of the paper is organized as follows. Section \ref{sec:system_design} provides an overview of DINAMITE design. Sections \ref{sec:log_format}, \ref{sec:inst_pass} and \ref{sec:inst_lib} contain a detailed discussion of the log format, LLVM instrumentation pass and logging library. Section \ref{sec:analysis_toolkit} discusses two implementations of analysis frameworks -- one in native C++ and another one using Spark Streaming. Section \ref{sec:evaluation} describes the tools we created with DINAMITE and evaluates them on three applications. Section \ref{sec:related_work} discusses related work. Section \ref{sec:future_work} elaborates on  possible avenues for future work.

%% file: systemdesign.tex
\section{System design} \label{sec:system_design}

\begin{figure}
    \centering
    \includegraphics[width=\columnwidth]{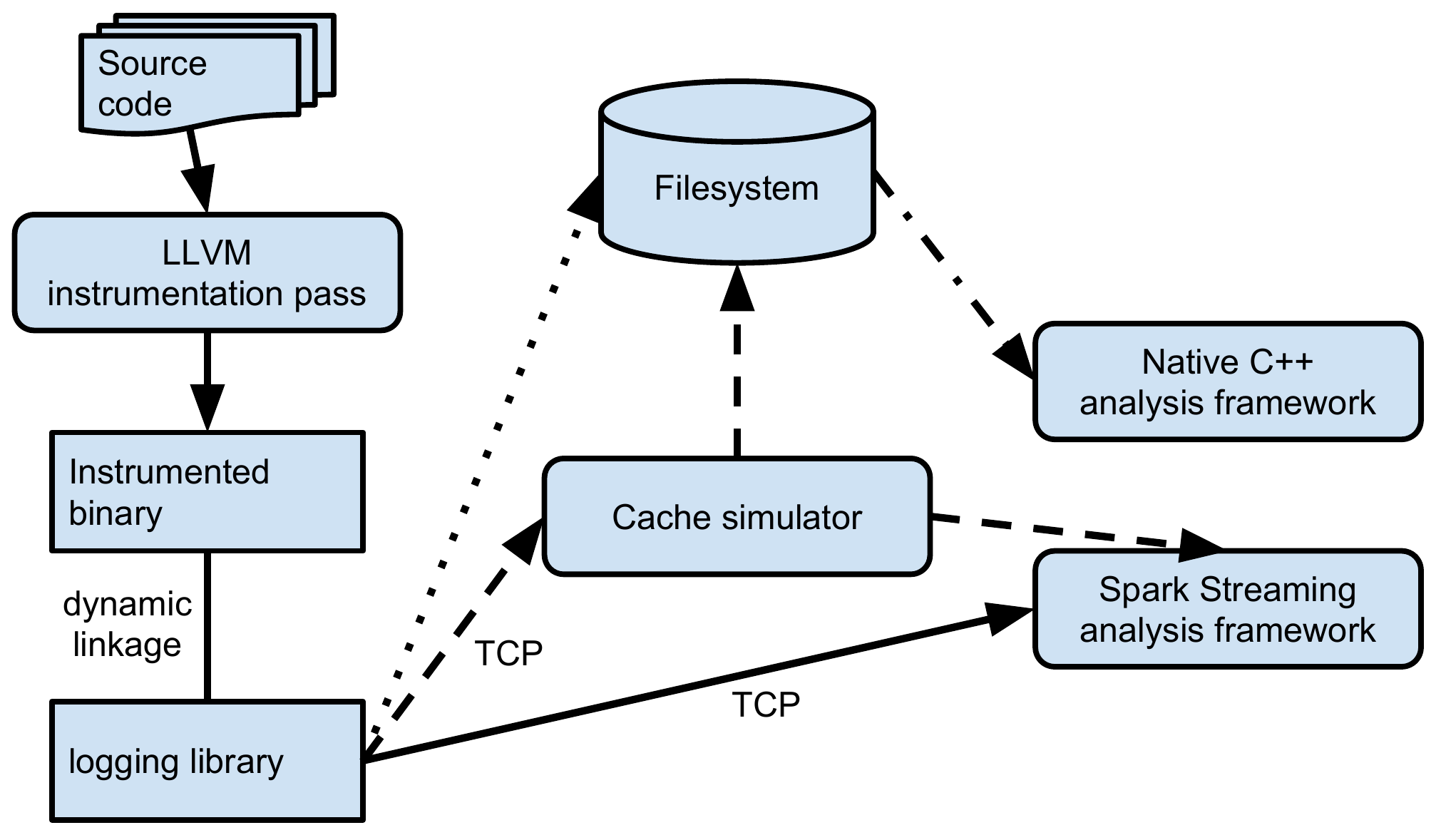}
    \vspace{-20pt}
    \caption{DINAMITE system diagram}
    \label{fig:inst_sys}
\end{figure}

Our system is built from three components: an LLVM instrumentation pass, a collection of output logging libraries and an analysis toolkit. A system overview is shown in Figure \ref{fig:inst_sys}. Different line types leaving the logging library show possible data paths within the system. A path taken by data depends on the type of analysis desired.

Target applications are compiled with an LLVM~\cite{lattner2004llvm} compiler configured to include our instrumentation pass. Configuration is trivial: it only requires changing the compiler invocation command. Most large software projects allow specifying the compiler command via an environment variable. 

Our instrumentation pass instruments three types of events: function entry or exit, memory allocation, and memory access. For each event the instrumentation pass injects a function call to an externally linked logging library. The logging library, linked dynamically at runtime, will produce a log record of an event in binary or text format (described below). Log records are either stored in the file system or streamed over a socket. Stored traces can be analyzed using pre-packaged DINAMITE scripts written in Python or C++ (see Section~\ref{sec:evaluation}), or the user can write her own tools using their language of choice. Log records streamed over the socket are processed by the Spark Streaming engine. DINAMITE includes several analysis kernels for Spark written in Scala; users can also write their own.

DINAMITE allows chaining analysis passes, similar to the Unix pipe command. For many of our analyses we stream the log data to a cache simulator tool (written in C++) that annotates each memory access log entry to indicate whether the access was a cache hit or a miss and forwards the  annotated log entry to the Spark Streaming engine. 

The rest of this section describes the system in more detail.

\input{llvmpass}
\input{logformat}

\input{logginglibs}
\input{analysistoolkit}

%% file: llvmpass.tex
\subsection{LLVM instrumentation pass} \label{sec:inst_pass}

We chose to use the LLVM infrastructure for the implementation for the following reasons:
\begin{enumerate}
    \item It can be used on programs written in any language supported by an LLVM front-end compiler. To date, those include, but are not limited to, C, C++, D, Haskell, Objective-C, Swift, Ruby, etc. There is even a compiler that translates Java bytecode into the LLVM intermediate representation. Given the popularity of LLVM, we can expect this list to grow in the future.
    \item It lets us add instrumentation at the level of the intermediate representation (IR), which is more convenient than instrumenting a binary. LLVM IR is an assembly-like language that is more abstract than machine code (e.g., it assumes an unlimited set of registers). When IR is translated into binary, a single memory access can be expanded into multiple machine instructions, which can introduce noise into traces and make it difficult to attribute accesses to source code level constructs. 
    \item Full debug information is available at IR level. The front end we used, \emph{clang}, embeds debug information into the IR as an abstraction of the DWARF format that is easy to parse with the tools provided in the LLVM framework.
\end{enumerate}

Our instrumentation pass begins by crawling the IR debug metadata to extract information about complex data types (structs, classes, unions) and categorizes them by connecting their corresponding internal LLVM references with type and field names. This is necessary because of type aliasing. A single type in C or C++ can have multiple names because of \verb|typedef|s. Without this metadata extraction, LLVM only knows about the original type definitions, but IR instructions may contain references to different names for the same type. We store all the type alias information in map-like data structures within the pass.

The core of the instrumentation pass iterates over the current module and visits each function, each basic block and each instruction within it.

For each encountered function, it places a \textit{function begin} log call at the beginning of its first basic block, and a \textit{function end} log call at the end of each basic block that ends the function.

Memory allocation functions are treated separately. Our pass must first recognize whether a function is a memory allocator and then gather the information about the allocated type, size and address. For each called function, our instrumentation checks the function's name against the list of known allocator functions. We generalize allocator functions as functions that take two arguments: \emph{number of elements} and \emph{size of a single element}, and output the address that points to the start of the allocated region. This model encompasses all allocation library APIs that we have encountered, and, combined with type information available through LLVM, contains all the relevant information that describes an allocation. 

The list of allocators is provided in a separate file, where each entry is described with a function name, and three indices indicating the position of all the relevant fields (the number of elements, the size of the element and the allocation's base pointer) in the argument list. If the allocation address is the function's return value, its index will be set to $-1$.  Standard allocation functions (such as \verb|malloc| and \verb|calloc|) are included in the configuration file that comes with our pass. If the program uses any non-standard allocator functions, the user must add them to that file. The pass places an allocation event log call after each call to an allocator function.

Strings written to the log are encoded with a unique integer identifier to preserve space. The integer-to-string mappings are placed into JSON documents created by the instrumentation pass. The mappings must be consistent across modules, but LLVM compiler passes do not preserve any state across modules. Therefore, we load the JSON mappings before compiling each module compilation and write back any updates at the end. 

For ease of use with C and C++ projects, our instrumentation pass gets registered with LLVM's pass manager for standalone clang invocations. As clang supports most of GCC's compilation flags, this makes integration of our instrumentation into existing projects in most cases as easy as changing the compiler invocation variable.

%% file: logformat.tex
\subsection{Log format}\label{sec:log_format}
All log events contain a field for a thread identifier. We limit this field to 8 bits to conserve space, but it can be easily expanded if needed. Distinguishing between 256 unique threads was sufficient for our case studies. 

Allocation and access events share fields that contain the file name, the line number and the column number that correspond to the event's source code location. Allocation events additionally contain the base address of the allocated memory region, the size of a single allocated element, the number of  allocated elements and their type. 

Access events contain the accessed address, the type of the access (read or write), the name and type of the variable corresponding to the access, and the value at the accessed address. 

Function events contain the event type (entry or exit into the function) and the function name.

Depending on the configuration, log records can be produced in text or binary format. In text format, each field of the entry is printed to a file, separated by a delimiter character. In binary format, different types of events are contained in a parent \verb|logentry| structure. The \verb|logentry| structure contains a type field that differentiates the payload as either a function, access or allocation event. The payload is a union between the corresponding three log entry types. In the current version of DINAMITE, each log entry takes 48 bytes total.

Log format involves a surprising trade-off between performance and log size, which we evaluate in the next section.

%% file: logginglibs.tex
\subsection{Logging libraries} \label{sec:inst_lib}

Instrumented programs do not contain any logic for producing log records. Instead, they contain calls to the externally linked logging library. Our implementation contains three different library versions based on log format and output destination: text-to-file, binary-to-file and binary-to-socket.

\noindent \textit{The effect of log format on log size}

In our implementation, each binary log record takes up 48 bytes of storage. Text entries are variable in size and depend on the number of characters needed to encode all the values. The two extremes of an entry size in text format are:
\begin{itemize}
    \item Minimum: 18 bytes. Each field can be encoded with a single digit, with added single character delimiters.
    \item Maximum: 77 bytes. Each field has the maximum value for its storage type in binary format.
\end{itemize} 

The reality is somewhere in-between, as shown in Table \ref{table:size_comparison}, which compares log sizes in text and binary format for SPEC2006 \emph{429.mcf}, with 4.5 billion memory accesses. Text format generates smaller logs; log size is important, because real workloads generate hundreds of gigabytes of logs. At the same time, using text format results in a much higher performance overhead (evaluated in the next section). Since we can forego storing large logs by relying on DINAMITE's streaming model we always use DINAMITE with the binary log format to avoid the overhead.

\begin{table}[]
\centering
\caption{Log size comparison in 429.mcf}
\label{table:size_comparison}
\begin{tabular}{|
>{\columncolor[HTML]{C0C0C0}}l |l|}
\hline
\textbf{Number of accesses:} & \cellcolor[HTML]{FFFFFF}$\sim$4.5 billion \\ \hline
\textbf{Binary log size:}    & 205GB                                    \\ \hline
\textbf{Text log size:}      & 172GB                                    \\ \hline
\end{tabular}
\end{table}

\noindent \textit{The effect of log format on performance}

For a detailed insight into the overhead of running the instrumented binary, we break it down into the following components:
\begin{itemize}
    \item \emph{base cost}: the cost of executing the uninstrumented code
    \item \emph{log cost}: the cost of invoking the logging library function
    \item \emph{format cost}: the cost of preparing the log entry for writing
    \item \emph{output cost}: the cost of writing the log entry
\end{itemize}

\begin{table}
    \centering
    \caption{Cost breakdown of text and binary formats for 429.mcf, per single log entry}    
    \label{tab:format_cost}
    \begin{tabular}{|
>{\columncolor[HTML]{C0C0C0}}l |l|l|}
\hline
\textbf{Format}      & \cellcolor[HTML]{C0C0C0}\textbf{Binary} & \cellcolor[HTML]{C0C0C0}\textbf{Text} \\ \hline
\textbf{Base}        & \multicolumn{2}{c|}{2.33 ns}                                                    \\ \hline
\textbf{Log cost}    & \multicolumn{2}{c|}{13.66 ns}                                                   \\ \hline
\textbf{Format cost} & 15.99 ns                                & 789.95 ns                             \\ \hline
\textbf{Total time (no output)}  & 31.98 ns                                & 805.94 ns                             \\ \hline
\end{tabular}

\end{table}

Table \ref{tab:format_cost} shows the broken-down cost of the instrumentation for \emph{429.mcf}. {We report all costs except the output cost, because it depends on where we write the log data; the output costs for different output destinations are reported later in this section.

Log cost is fixed and must be invoked for every instrumented event. The only way to reduce this cost is to avoid instrumenting certain events altogether, according to a user-defined criterion at compile time. For example, if we are not interested in exploring the entire memory access trace of a program, but only accesses to a single type, we can tell the compiler to instrument only those. Similarly, we can limit instrumentation to certain functions. 
Instrumenting isolated data structures or types can discover some memory access patterns, but this kind of filtering is not appropriate for understanding whole program memory behavior or for any analysis involving cache simulation.

Format cost is the cost of packing the log data according to the specified format. It is dominant for the text format, because formatting strings is a very expensive operation, relative to producing a binary record. Text format suffers a 25$\times$ higher run time relative to the binary format. That is why we always resort to using the binary format in our experiments, despite its higher storage overhead.

\begin{figure}
    \centering
    \includegraphics[width=\columnwidth]{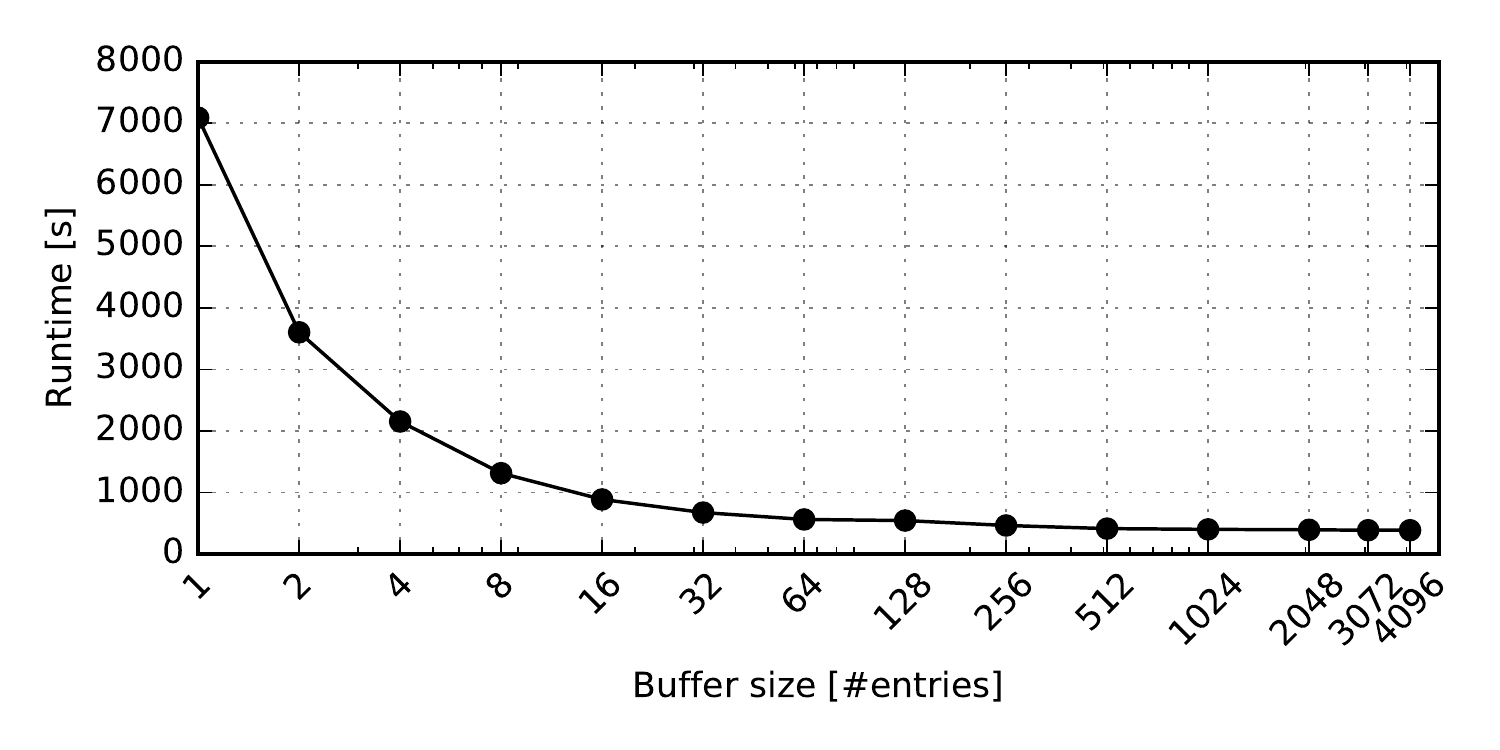}
    \vspace{-25pt}
    \caption{Impact of buffering on performance of 429.mcf}
    \label{fig:buffering}
\end{figure}

Output cost is the cost of writing the log records into a file or sending them over a socket. Besides the cost of accessing the storage medium, it requires a system call. We mitigate this overhead via buffering. Figure \ref{fig:buffering} shows the effects of different  buffer sizes on the runtime of 429.mcf in the binary format that writes to a RAM disk. By increasing the output buffer size, performance improvement reaches its maximum at around 20$\times$ over the unbuffered version. In the rest of our measurements, we used the output buffer of 4096 entries.

Table \ref{fig:overhead_comparison} compares the performance of \emph{429.mcf} instrumented with  DINAMITE against slowdowns of two major instrumentation frameworks: Intel Pin and Valgrind.
Valgrind performance degradation reported here is obtained from Nethercote et al. \cite{nethercote2007valgrind}, and refers to Valgrind's MemCheck tool which performs memory error checking with a summary output at the end of execution. This is not a fair comparison to access instrumentation, but to the best of our knowledge, a Valgrind tool comparable in functionality with DINAMITE is not available. Numbers for Pin were obtained from the supplied \verb|pinatrace| tool, output to RAM disk. In Table \ref{fig:overhead_comparison}, the slowest  version  of  DINAMITE  without  analysis  instruments each access with full debug information (as described in section \ref{sec:log_format}) and outputs it to a RAM disk filesystem in binary format. Even at this level of detail and with full output enabled, DINAMITE is only 60\% slower than Valgrind's MemCheck and almost 10x faster than the comparable access instrumentaion in Pin. Even when using the Spark analysis pipeline, DINAMITE is only 35\% slower than \verb|pinatrace|.


Table \ref{fig:output_costs} compares the running times for executing  \emph{429.mcf} with different variants of log formats and outputs. Note that the text-formatted output makes the instrumentation run very slow: 33$\times$ slower than using the binary format. Sending the trace over a TCP socket to \verb|netcat| is faster than writing it to a RAM disk. However, introducing Spark Streaming into the pipeline makes the TCP streaming execution 15x slower. Optimizing this would require a detailed analysis of Spark's data receiving system and is left for future work.

\begin{table}[width=\linewidth]
\centering
\caption{Instrumentation overhead comparison - 429.mcf}
\label{fig:overhead_comparison}
\begin{tabular}{|l|l|}
\hline
\rowcolor[HTML]{C0C0C0} 
\textbf{Framework}                           & \textbf{Slowdown} \\ \hline
\textbf{Pin (pinatrace output to RAM disk)}            & 354x                       \\ \hline
\textbf{Valgrind (MemCheck)}                            & 22x                              \\ \hline
\textbf{DINAMITE (empty instrumentation)}    & 7x                             \\ \hline
\textbf{DINAMITE (binary format, no output)} & 14x                             \\ \hline
\textbf{DINAMITE (binary format, output to RAM disk)}       & 36x                            \\ \hline
\textbf{DINAMITE (Spark analysis)}    & 537x                             \\ \hline
\end{tabular}
\end{table}

Our design decouples the generation of log records from their processing. Alternatively, embedding analysis logic into the logging library is also possible. We opted against it for the following reasons: 
\begin{itemize}
    \item Instrumented programs share heap with the logging library. Adding significant bookkeeping data structures to the heap could affect the placement of the program's data and diminish the accuracy of traces.
    \item Decoupling analysis from logging allows for flexibility in the languages and frameworks used for analyzing memory traces 
\end{itemize}

\begin{table}[width=\linewidth]
\centering
\caption{Logging library performance - 429.mcf}
\label{fig:output_costs}
\begin{tabular}{|l|l|l|l|}
\hline
\rowcolor[HTML]{C0C0C0} 
\textbf{Version}         & \textbf{Destination} & \textbf{Time {[}s{]}} & \textbf{Slowdown} \\ \hline
Uninstrumented           & nil                  & 10.05                 & 1x                \\ \hline
Text (unbuffered)        & RAM disk             & 11820                 & 1176x             \\ \hline
Binary (file) (buff.) & RAM disk             & 360.09                & 36x               \\ \hline
Binary (file) (buff.) & Hard disk            & 1426                  & 142x              \\ \hline
TCP (buffered)           & netcat \textgreater  /dev/null               & 339.12                & 34x               \\ \hline
TCP (buffered)           & Spark (access count)          & 5400                  & 537x              \\ \hline
\end{tabular}
\end{table}

%% file: analysistoolkit.tex
\subsection{Analysis toolkit}\label{sec:analysis_toolkit}

The analysis toolkit consists of two different frameworks for writing log processing applications. Logs recorded to a filesystem are processed with the native analysis framework written in C++. The framework includes support for parsing logs and allows the user to easily extend the analysis by writing a new C++ class. Alternatively, the users could write their own parsing and analysis tools in any language of choice. Logs streamed over a TCP socket are processed live with Spark Streaming drivers. Similarly, the user could configure DINAMITE to use any another system to ingest or analyze streaming logs (e.g., Kafka, Google Dataflow). Our toolkit includes a simple cache simulator program, which processes and annotates streamed log records with cache hit/miss indicators. Next, we describe these components in more detail.

\input{nativefw}
\input{sparkfw}
\input{cachesim}

%% file: nativefw.tex
\subsubsection{Native analysis framework}

The C++ framework provides support for writing arbitrary analysis kernels. To write a new kernel, the programmer must extend the \verb|TracePlugin| class (shown in listing \ref{lst:cpp_example}). 

\begin{lstlisting}[style=style1, language=C++, caption=Trace plugin base class, label=lst:cpp_example]
class TracePlugin {
...
 protected:
 NameMaps *nmaps;
 TracePlugin(const char *name);
 public:
 virtual void processLog(logentry *log) =0;
 virtual void finalize() =0;
 virtual void passArgs(char *args) =0;
};
\end{lstlisting}

The framework reads log records into a buffer and passes each log entry to the chosen plugin by invoking its \verb|processLog(logentry*)| method.  At the end of the log file, the framework calls the plugin's \verb|finalize()| method, which is used for writing the output of the analysis.

%% file: sparkfw.tex
\subsubsection{Spark Streaming analysis framework}

To analyze streamed log records with Spark Streaming, the programmer must write an analysis kernel in  Scala. To this end, our framework provides a custom Spark Streaming \verb|Receiver| class and a log converter. A receiver accepts batches of log events in binary format over a TCP socket and stores each separate log entry in its associated \verb|StreamingContext|.

Listing \ref{lst:spark_example} shows a Spark Streaming kernel for counting the number of memory accesses per variable. To get useful information out of the entries, the incoming \verb|DStream| is routed through a \verb|map| operation which invokes our \verb|LogConverter| class on each separate entry. \verb|LogConverter| unpacks and outputs log data as Scala classes, with the distinction between function events, allocation events and access events. To get a \verb|DStream| of instances of a certain event type, logs are filtered with a class matching operation. These events are then mapped to \verb|(varId, 1)| pairs, and reduced by summing over variable IDs. Persistent state is updated by invoking Spark Streaming's \verb|updateStateByKey()| operation. A custom update function, omitted in our listing for brevity, updates the counts by summing new results with the previous state. Results are then output to the console or the filesystem.

\begin{lstlisting}[style=style1,language=Scala, caption=Example Spark Streaming kernel, label=lst:spark_example]
def main(args: Array[String]) {
  val sparkConf = new SparkConf()
    .setAppName("AccessCounter");
  val ssc = new StreamingContext(sparkConf,
    new Duration(1000));
  ssc.checkpoint("/checkpoints/");

  val logs = ssc
    .receiverStream(new LogReceiver(9999))
    .map(rawlog => 
    LogEntryReader.extractEntry(rawlog));

  val counts = logs
    .filter(log => 
      log.isInstanceOf[AccessLog])
    .map(access =>
    (access.as(...)[AccessLog].varId, 1L))
    .reduceByKey(_+_)
    .updateStateByKey(sumUpdater);

  counts.print();

  ssc.start();
  ssc.awaitTermination();
}
\end{lstlisting}

Integration with Spark Streaming gives the programmer access to the full set of Spark Streaming operations and can process logs as they are output from a live running program.

%% file: cachesim.tex
\subsubsection{Cache simulator}

For detailed analysis of program cache behaviour, we wrote a simple cache simulator, which is placed as an intermediate step between the generation of the log output and the analysis framework (or the filesystem, if we are saving the logs for offline processing).

We simulate a single-level cache, typically configured with parameters reflecting a last-level cache on our target system. The cache simulator accepts log entries over a socket, much like the Spark Streaming analysis framework. It annotates each memory access with an indicator whether this was a cache hit or a miss. The annotated logs are passed on to either the analysis framework, or stored in a filesystem for offline processing.

In our evaluation of the system, we found that having cache behaviour information was essential for identifying certain optimization opportunities.

%% file: evaluation.tex
\section{Evaluation} \label{sec:evaluation}




In this section we describe three tools that we built using DINAMITE and demonstrate how they guided our optimization of three applications: SPEC's \textit{429.mcf}, PARSEC's \textit{fluidanimate} and \textit{WiredTiger}, a MongoDB's  key-value store.

All experiments described in this section were performed on machines listed in Appendix A.

\subsection{Identifying cache offenders}

\epigraph{"For large working sets it is important to use the available
cache as well as possible. To achieve this, it might be
necessary to rearrange data structures. While it is easier
for the programmer to put all the data which conceptually
belongs together in the same data structure, this might
not be the best approach for maximum performance."}{Ulrich Drepper, 2007\cite{drepper2007every}}

This tool identifies program variables and source lines that generated the most last-level cache accesses and misses. It works as follows:

\vspace{10pt}
\noindent \textit{429.mcf}

\textit{429.mcf} performs single-depot vehicle scheduling using the network simplex method. The implementation represents nodes and arcs in the network as C \verb|struct|s. In the benchmark description the author mentions reordering fields of both node and arc structs in an attempt to reduce cache misses and improve performance \cite{1_henning_2006}. Nevertheless, DINAMITE enabled additional optimizations. 

\begin{table}[]
\centering
\caption{429.mcf top miss offenders}
\label{tab:mcf_misses}
\begin{tabular}{|l|l|l|l|}
\hline
\rowcolor[HTML]{C0C0C0} 
\textbf{Variable Name} & \textbf{File} & \textbf{Line} & \textbf{Miss count} \\ \hline
arc.ident              & pbeampp.c     & 167           & 45247271            \\ \hline
node.orientation       & mcfutil.c     & 85            & 1104784             \\ \hline
node.basic\_arc        & mcfutil.c     & 86            & 988543              \\ \hline
arc.cost               & mcfutil.c     & 86            & 767273              \\ \hline
node.potential         & pbeampp.c     & 170           & 235696              \\ \hline
\end{tabular}
\end{table}

Table \ref{tab:mcf_misses} shows the output of the cache-offender tool. We notice that a disproportionate number of misses are being caused by the \verb|ident| field of the \verb|arc| struct, more than four times as many as the second most accessed field, \verb|node.orientation|.

Upon closer inspection, we noticed that all the \verb|arc| structures are allocated as a single large array, even though they represent nodes in a linked data structure. The majority of accesses to \verb|arc.ident| were made within a single loop (shown in Listing \ref{lst:mcf_traversal}). The loop iterates over the \verb|arc| array until it finds a match, and only then accesses its other fields.

Every time \verb|arc.ident| was accessed, the corresponding cache line was filled with other fields, most of which were not used before the cache line was evicted. \conor{Possible rewrite (removes article "these", and uses a more active tone to make a more bold statement about how this is a problem): Access patterns to data layouts like these waste cache space and memory bandwidth.} These data layout and access pattern waste cache space and memory bandwidth. We addressed the  problem by restructuring the array of \verb|arc|s from the \textit{array of structures} layout into the \textit{structure of arrays}.

\begin{lstlisting}[style=style1, firstnumber=165, language=C, label=lst:mcf_traversal, caption={429.mcf pbeampp.c excerpt}]
for( ; arc < stop_arcs; arc += nr_group )
  {
    if( arc->ident > BASIC )
      {
        /*red_cost = bea_compute_red_cost(arc);*/
        red_cost = arc->cost - arc->tail->potential + arc->head->potential;
        if( bea_is_dual_infeasible( arc, red_cost ) )
          {
            basket_size++;
            perm[basket_size]->a = arc;
            perm[basket_size]->cost = red_cost;
            perm[basket_size]->abs_cost = ABS(red_cost);
          }
      }
  }

\end{lstlisting}

Our modifications brought a 55\% reduction in LLC misses, and a 12\% improvement in the overall runtime. 

\vspace{10pt}
\noindent \textit{fluidanimate}

Fluidanimate is an Intel \textit{Recognition, Mining and Synthesis} application that uses the Smoothed Particle Hydrodynamics method to simulate an incompressible fluid. It uses the Navier-Stokes equation to derive fluid density fields. It is included in the PARSEC 3.0 benchmark suite because of the increasing significance of physics simulation in video-game programming and real-time animation domains~\cite{bienia11benchmarking}.

Profiling \textit{fluidanimate} with \verb|perf| \cite{de2010new} showed that it has a high LLC miss rate of 30\% on our system. We instrumented the program using DINAMITE and ran it through the cache offender tool. Table \ref{tab:fluidanimate_misses} shows output of the tool: variable names and source lines responsible for the most cache misses. The top cache offenders are \verb|Cell.next| and \verb|Vec3.x|. The names of the structs and fields suggest that a \verb|Cell| is an element of a linked collection. Listing \ref{lst:fluidanimate_traversal} shows the code excerpt pointed to by the output of our tool. We can immediately see that the code generating misses is a traversal of grid of \verb|Cell| structures in which only the \verb|next| field is touched.

Looking at the definitions for \verb|Cell| and \verb|Vec3| types we can see that \verb|Cell| represents a linked list of containers for arrays of \verb|Vec3| structures that contain three-dimensional vectors. The arrays themselves are contained within the \verb|Cell| struct in their entirety. The total size of a \verb|Cell| struct with the payload was 896 bytes, making a single instance span 14 cache lines.

\begin{table}[]
\centering
\caption{CSV output of the miss summary tool for\\fluidanimate}
\label{tab:fluidanimate_misses}
\begin{tabular}{|l|l|l|l|}
\hline
\rowcolor[HTML]{C0C0C0} 
\textbf{Variable name} & \textbf{File} & \textbf{Line} & \textbf{Miss count} \\ \hline
Cell.next              & pthreads.cpp  & 530           & 184496              \\ \hline
Vec3.x                 & ./fluid.hpp   & 354           & 95682               \\ \hline
Cell.next              & ./fluid.hpp   & 404           & 73800               \\ \hline
Vec3.x                 & ./fluid.hpp   & 346           & 67327               \\ \hline
Vec3.x                 & ./fluid.hpp   & 355           & 66657               \\ \hline
\end{tabular}
\end{table}

This data layout is poorly optimized for traversing lists of \verb|Cell|s, because each new \verb|Cell| access generates a cache miss. Our idea, therefore, was to allocate the \verb|Cell|'s payload, which is rarely touched, separately from the rest of the structure. The structure would then include a pointer to its payload; since each \verb|Cell|'s payload consists of multiple arrays, adding a layer of indirection to access the payload  would not be much of a penalty. Allocating the \verb|Cell| payload separately brings the size of the structure down to 16 bytes . Since consecutive calls to a memory allocator function for a variable of the same size will return near consecutive addresses in most standard libraries, consecutive \verb|Cell|s will be allocated close together,  and several of them will fit into a single cache line.

\begin{lstlisting}[style=style1, firstnumber=522, language=C, label=lst:fluidanimate_traversal, caption={fluidanimate pthreads.cpp code excerpt}]
void ClearParticlesMT(int tid)
{
  for(int iz = grids[tid].sz; iz < grids[tid].ez; ++iz)
    for(int iy = grids[tid].sy; iy < grids[tid].ey; ++iy)
      for(int ix = grids[tid].sx; ix < grids[tid].ex; ++ix)
        {
         int index = (iz*ny + iy)*nx + ix;
         cnumPars[index] = 0;
         cells[index].next = NULL;       
         last_cells[index] = &cells[index];
        }
 }
\end{lstlisting}

\begin{figure}
    \centering
    \includegraphics[width=\columnwidth]{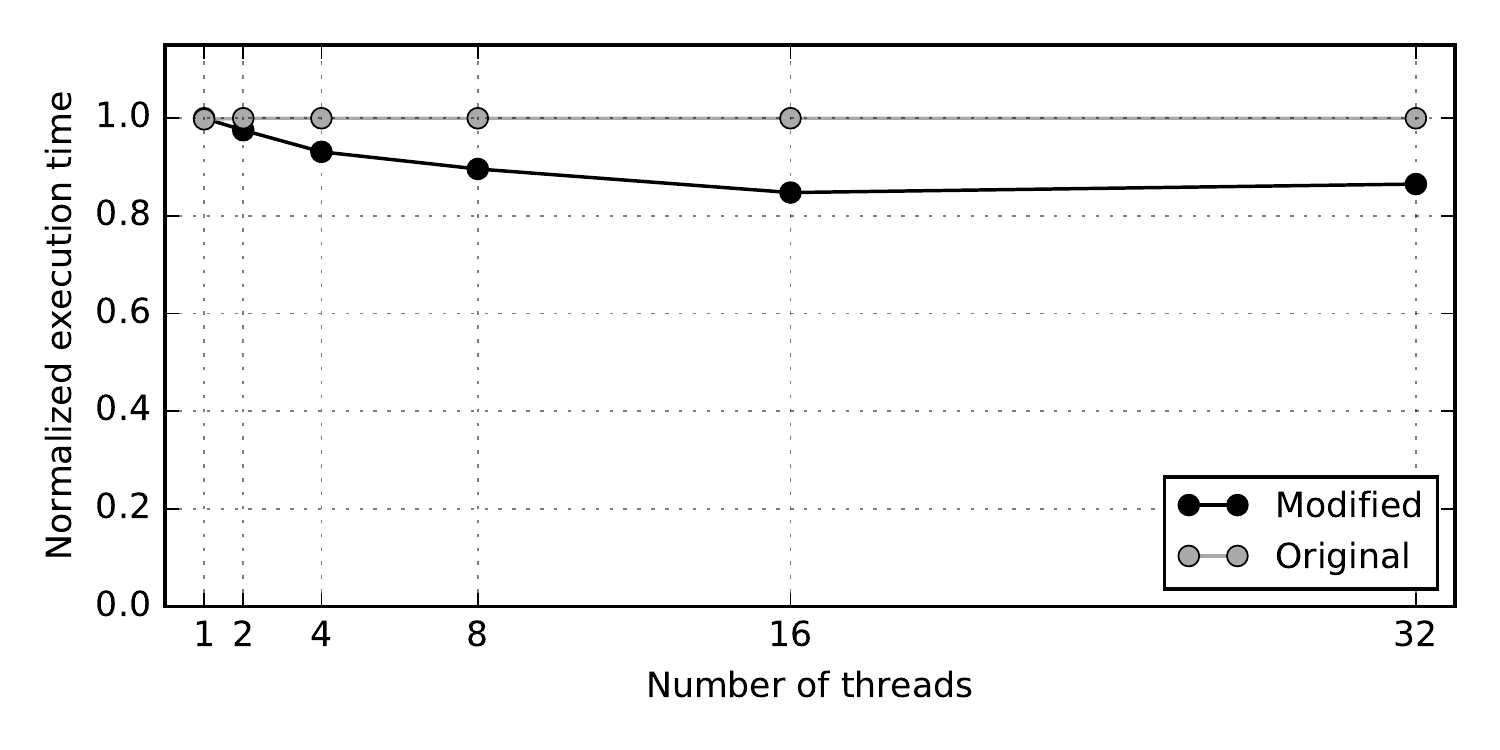}
    \vspace{-20pt}
    \caption{Scaling improvements in PARSEC3.0 fluidanimate application}
    \label{fig:fluidanimate_results}
\end{figure}

This change brought a 50\% reduction in the LLC cache miss rate and a 15\% reduction in runtime with 16 threads (see Figure \ref{fig:fluidanimate_results}). 

An interesting observation is that \verb|Cell|s were  allocated in the original implementation to be cache-aligned and padded to a fill the entire cache line, indicating a prior effort to make better use of the cache hierarchy. However, with DINAMITE we discovered that making the \verb|Cell| struct larger by padding actually hurt performance on our system. Intricacies of modern multi-core memory hierarchies can mislead even very experienced programmers. Powerful performance analysis tools are thus crucially important.

\subsection{Structure splitting}
Our structure splitting tool is based on the class splitting algorithm proposed by Chilimbi et al. for Java programs~\cite{chilimbi1999cache}. The algorithm analyses how the program accesses the members of a class to determine if a class is fit to split into two separate classes. Splitting classes is motivated by the idea that \textit{hot} fields, or fields that are accessed significantly more than \textit{cold} fields, should be placed in a separate class so that more hot data can be packed into a single cache block. To access fields that are considered cold, the hot class includes a pointer to the cold class. (This is similar to the optimization that we applied to \textit{fluidanimate}). 

The algorithm begins by identifying \textit{live} classes. A class is considered \textit{live} if it is accessed more than a certain threshold, and only live classes are considered for splitting. Next, fields in live classes are marked as hot or cold depending on how many times their respective class is accessed. If a field is accessed significantly more than other fields, it is considered hot. Full details of the algorithm are described elsewhere~\cite{chilimbi1999cache}.

Chilimbi et al. implemented the splitting algorithm for Java classes, using the JVM for access statistics and a Java byte-code instrumentation tool BIT \cite{lee1997bit}. We implemented the splitting algorithm in DINAMITE, making it accessible to a wider range of programs, including those written in unmanaged languages. The tool works as follows: 

The Spark Streaming driver receives access logs from the instrumented binary and produces the list of variables and corresponding access counts. Then a Python script generates a chart for each live structure showing weights assigned by the algorithm for individual fields; black bars for hot fields and gray bars for cold fields. Programmers then split their structures according to the hot and cold fields in the chart.



\begin{figure}
    \begin{minipage}{0.65\columnwidth}
    \subfloat{
        \includegraphics[width=\columnwidth]{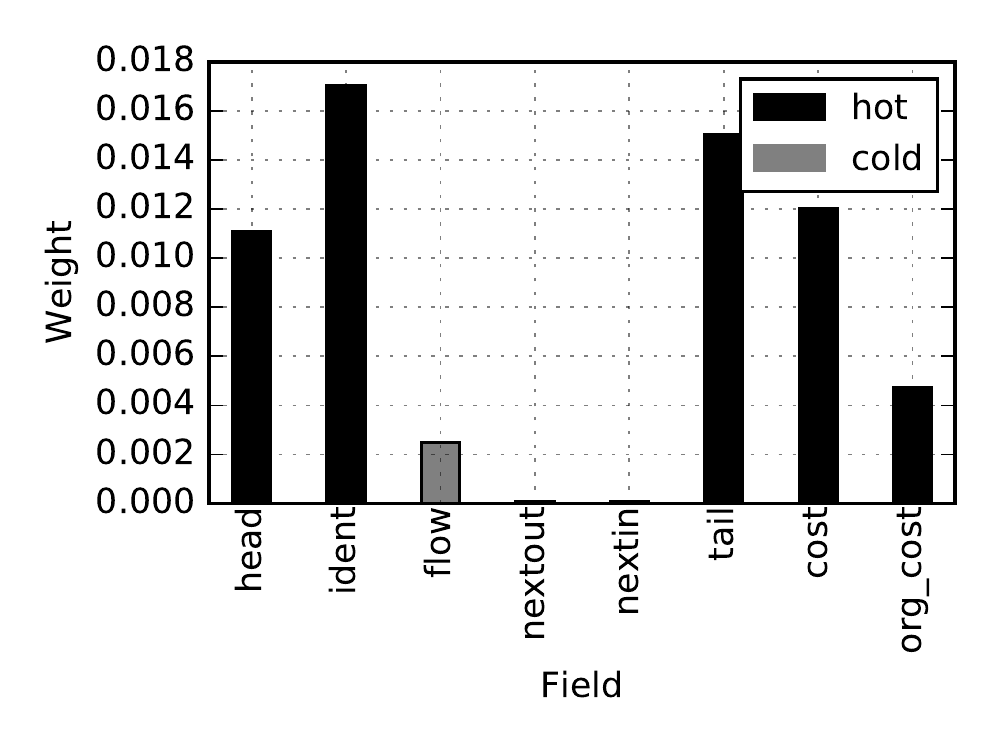} 
        }
    \end{minipage}%
    \hfill
    \begin{minipage}{0.3\columnwidth}
    \vfill
    \subfloat{
\begin{lstlisting}[basicstyle=\tiny]
struct cold_arc {
  flow_t flow;
  arc_p nextout;
  arc_p nextin;
};

struct hot_arc {
  cost_t cost;
  node_p tail, head;
  int ident;
  cost_t org_cost;
  struct cold_arc *ca;
};
\end{lstlisting}
    }
    \end{minipage}
    \vspace{-10pt}
    \caption{Tool output and modified code for structure splitting of 429.mcf}
    \label{fig:splitting}
\end{figure}

Figure~\ref{fig:splitting} shows the chart produced by the structure splitting tool for the \verb|arc| struct in \textit{429.mcf} as well as the modified \verb|arc| struct code. Similarly struct \verb|node| (not shown) was another live struct with both hot and cold fields. Splitting hot and cold fields in these structs delivered  20\% speedup and reduced the LLC miss rate by 60\%, as measured with \verb|perf|.

\subsection{Shared variable detection} \label{sec:shared_var}

On machines with even a handful of cores, variables updated by multiple threads can quickly become a scaling bottleneck~\cite{boyd2010analysis}, even if these variables are not protected by a lock or accessed via atomic instructions~\cite{dice2013scalable}. Repeatedly updating a shared variable from different cores  stresses the coherency protocol and can slow down the program by an order of magnitude relative to a sharing-free execution. Tools for detecting shared variable bottlenecks do exist, but they are hardware-specific (e.g., Intel's VTune~\cite{malladi2009using} works only on Intel machines, DProf~\cite{pesterev2010locating} and Memprof~\cite{lachaize2012memprof} work only on AMD hardware) and can be non-trivial to set up (DProf and Memprof require changing the kernel). DINAMITE is easily extended to detect shared variable bottlenecks on any binary that can be compiled with LLVM. 

To demonstrate, we created a simple tool that we then used to \svetozar{quickly} find a known scalability bottleneck in WiredTiger~\cite{wt}  \cite{wt-pull-request}, a MongoDB storage engine~\cite{wt-mongo-db}. To truly test the experience of creating new tools for DINAMITE, the student who created the shared variable detection tool was \textit{not} informed what variables and source locations triggered the bottleneck; he was only advised that the bottleneck exists and provided the instructions for running the problematic workload. 

The engineer who originally diagnosed the scalability issue took about week to do so after observing poor performance; she used Memprof, which required communication with its authors and changes to the kernel. Even though the changes were simple, they would likely be considered ``beyond the call of duty'' by many developers.  The DINAMITE tool took several hours to create by a person familiar with the overall framework and consisted of two simple Spark Streaming kernels and a Python script. 

The first and second kernel identify top shared variables. The second kernel processes the execution traces again, looking only for frequently shared variables and collects the source locations where the accesses are made. The tool could be structured with only a single kernel that both identifies the top shared variables and records the source lines, but we found that having two kernels is simpler and results in better performance of Spark Streaming.

The first kernel translates memory access log entries into  (\emph{accessed address}, \emph{variable identifier} and \emph{thread identifier}) tuples. Each tuple acts as the key in \emph{map-reduce} transformation that produces a list of variable-identifying tuples and the corresponding total memory accesses. The result is stored in a persistent table. 

Next, a Python script reads that table and transforms the data into a dictionary, where each accessed address serves as a key and the corresponding value contains the variable identifier and access counts performed by each thread.
The script filters the results according to the following criteria:
\begin{itemize}
    \item It removes all entries accessed by only a single thread
    \item It removes all entries that are not uniformly shared  by threads. We define uniform sharing as follows:
    
    Let $A_{sorted}$ be a list of all the per-thread access counts for the address, sorted in descending order, zero indexed.
    
    if $A_{sorted}[0] < 2 * A_{sorted}[1]$ the sharing is uniform.
    
\end{itemize}

The output is then sorted in descending order by the total number of accesses. Table \ref{tab:wt_accesses} shows the first five entries of the output generated for the LevelDB sequential read benchmark over WiredTiger (release 2.6.0) executed with 32 threads\footnote{The connection structure is shown as accessed by more than 70 threads because the benchmark creates and tears down additional threads before the measured run.}, which triggered the bottleneck. The top offender is the field \verb|v| in \verb|__wt_stats| struct. 

These results only point to the variable responsible for shared accesses. To find the root cause, we use the second Spark kernel to find the source location where the accesses are performed. The second kernel is very similar to the first, except it discards all the log entries that do not correspond to the top shared variable, and keeps the source lines where the accesses occurred. We sort the results by the number of accesses in descending order. 

\begin{lstlisting}[style=style1, label=lst:wt_finalresult, caption={WiredTiger shared variable analysis result (JSON)}]
{
  "threadcount": 18, 
  "totalcount": 311881, 
  "threads": [
   [
    156, 
    19492
   ], 
   [
    163, 
    19520
   ],
   ...
   ], 
  "file": "wiredtiger/build_posix/../src/btree/bt_curnext.c", 
  "line": 446, 
  "variable": "__wt_stats.v"
 }

\end{lstlisting}

Listing~\ref{lst:wt_finalresult} shows the first entry of the output (redacted for brevity), which correctly identifies the source location responsible for the bottleneck. The fields in the output JSON document are self-explanatory apart from the \verb|threads| list, which contains \verb|(thread id, thread access count)| pairs. It turns out that this sequential read-only workload suffered from scalability problems, because threads were incrementing a shared statistics counter after each read operation. This problem was later fixed by implementing per-thread statistic buffers.

\begin{table}[]
\centering
\caption{Most accessed shared variables}
\label{tab:wt_accesses}
\begin{tabular}{|l|l|l|l|}
\hline
\rowcolor[HTML]{C0C0C0} 
\textbf{Address} & \textbf{\# Accesses} & \textbf{\# Threads} & \textbf{Variable}                \\ \hline
\textit{0x64D900} & \textit{42495568}     & \textit{32}           & \textit{\_\_wt\_stats.v}         \\ \hline
0x64D1A4          & 26183326              & 74                    & \_\_wt\_connection\_impl         \\ \hline
0x64E0EC          & 7233836               & 72                    & \_\_wt\_connection\_impl.N/A \\ \hline
0x64D100          & 4786616               & 36                    & \_\_wt\_txn\_global.states       \\ \hline
0x64D540          & 4786370               & 34                    & \_\_wt\_stats.v                  \\ \hline
\end{tabular}
\end{table}

Figure \ref{fig:wt_results} shows the performance impact of the bug on Machine A (described in the appendix).  A seemingly benign counter increment, which takes negligible time in a single-threaded execution, quickly escalates into a huge scaling bottleneck with as few as four threads and slows down the workload by an impressive factor of 20 with 32 threads.  Previous work reported similar performance impact of shared variables on multicore machines~\cite{dice2013scalable}. With the increasing core counts on new hardware the importance of tools that enable productive memory performance analysis will continue to grow.

\begin{figure}
    \centering
    \includegraphics[width=\columnwidth]{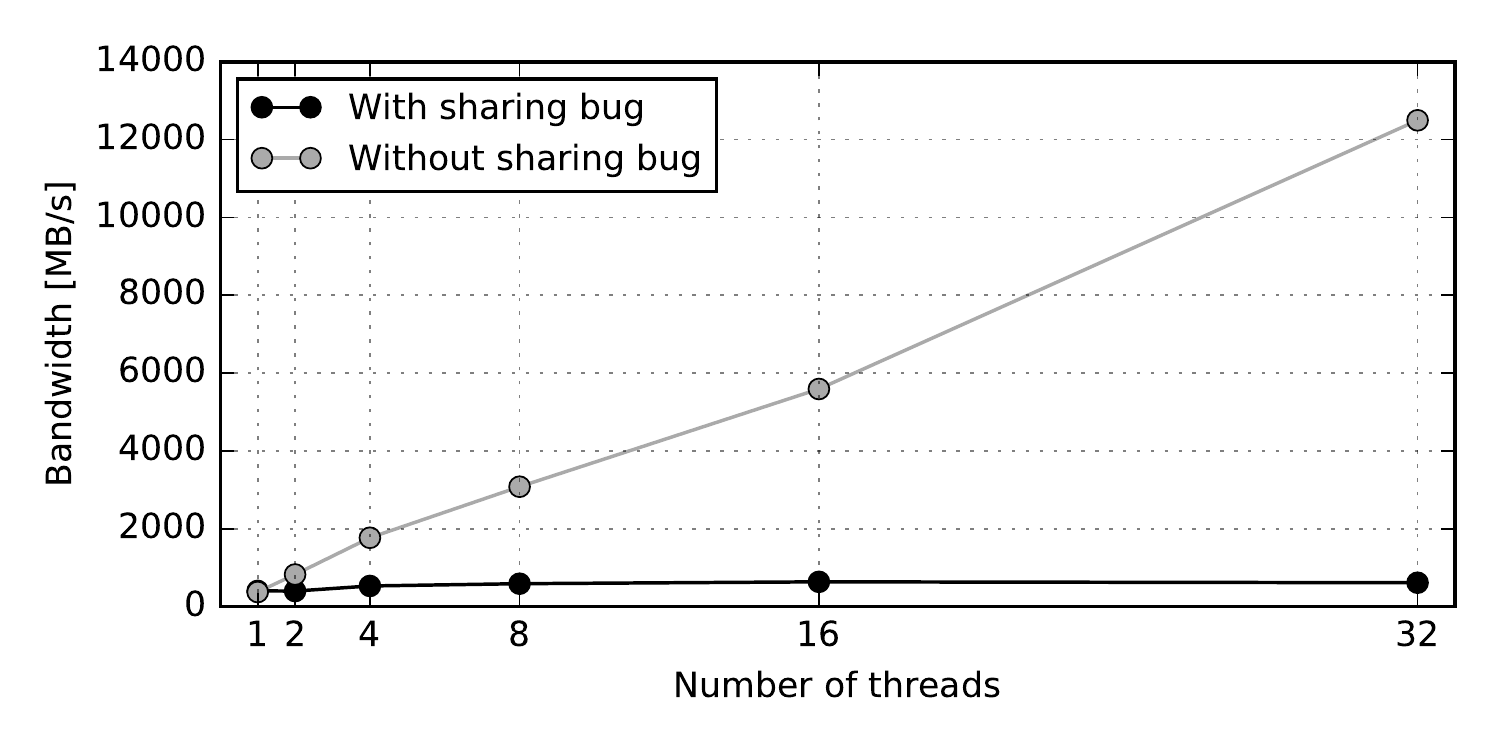}
    \vspace{-20pt}
    \caption{Scaling improvements WiredTiger after removing the shared variable bug}
    \label{fig:wt_results}
\end{figure}


%% file: relatedwork.tex
\section{Related Work} \label{sec:related_work}
        
Several instrumentation frameworks have been designed previously with the goal of solving memory bottleneck problems. Limitations of these frameworks include targeting a specific programming language, providing coarse grained instrumentation abstractions, or providing fine grained instrumentation abstractions that are sampled to reduce overhead. Other tools are not as flexible in that users can only view output through a user interface, or are designed for a specific use case.

Existing instrumentation frameworks like Pin \cite{luk2005pin} and Valgrind \cite{nethercote2007valgrind} instrument programs and allow programmers to build dynamic analysis tools on top of them much like DINAMITE. Pin instruments using a highly optimized JIT compiler that intercepts the first instruction of a supplied executable and compiles instrumentation functions to execute where appropriate. Valgrind recompiles code blocks from the binary into its own IR, instruments them using a tool plugin, and compiles them back to machine code to be executed. Additionally, Valgrind provides a fast shadow memory implementation for use in the tool plugins. Both of these frameworks have access only to information contained within the executable, making it difficult to relate machine instructions to source-level context.  DINAMITE instruments programs at compile time, at which point all the source-level information is available and automatically passed to the logging functions. The need for shadow memory is eliminated by decoupling DINAMITE's instrumentation from its analysis frameworks.

\svetozar{
Existing instrumentation frameworks like Pin \cite{luk2005pin} and Valgrind \cite{nethercote2007valgrind} instrument programs and allow programmers to build dynamic analysis tools on top of them much like DINAMITE. Pin achieves instrumentation through a highly optimized JIT compiler that intercepts the first instruction of a supplied executable and compiles instrumentation functions to execute where appropriate. Pintools are created by the programmer using the Pin framework that describe the instrumentation analysis. Runtime binary instrumentation is limited in that information gleaned from the framework does not translate to code level data structures out of the box. Without knowing what machine instructions relate to higher level data, programmers do not immediately see patterns related to their original work. DINAMITE instrumentation is inserted at compile-time allowing it to retain code level information about data that is valuable to the programmer using the tool. 
}
\svetozar{
Valgrind is another framework that performs runtime binary instrumentation. It addresses the lack of code level information other runtime analysis frameworks suffer from by supporting a technique called "shadow values". Shadow values replace values in memory and registers with values that describe them. Valgrind requires tool writers to implement their own shadow values -- a technique difficult to implement. The framework supports the implementation of shadow values by providing registers to store shadow values and extra output functionality for printing these values during execution. Though this technique enriches instrumentation by providing more context about memory accesses during execution, it is a complex task in practice, and can easily slow down instrumentation if not done carefully. DINAMITE provides memory access information by default; each log entry contains the address of the memory location accessed and is available for tool writers to query, though register information is absent but can be easily extended using the llvm.read\_register instrinsic \cite{llvm-read-reg}. Programmers using DINAMITE are not left to implement complex and bug-prone functionality to obtain memory access information.
}

Zhao et al. \cite{zhao2011dynamic} describe a tool designed to detect true and false sharing built on top of the memory shadowing framework Umbra \cite{zhao2010umbra}. \svetozar{Umbra is a performant implementation of shadow values while remaining general purpose, and is achieved by mapping only allocated memory instead of the whole address space. } Memory sharing is detected by associating cache lines with shadow memory exposed by Umbra. Association of thread to address is done via a bitmap describing thread ids responsible for each access. Sheriff \cite{liu2011sheriff} addresses the same problem, but requires either the programmer to rewrite source code or rely on sampling to catch culprits, and is specifically designed to detect false sharing. DINAMITE achieves the same result by inserting thread IDs in each log entry, which also contain accessed addresses. The log entry also contains all the source level details necessary to pinpoint exactly where in the code the accesses were performed and to what data type.

Memprof \cite{lachaize2012memprof} is a tool that profiles objects that make remote memory accesses on NUMA machines so programmers can potentially minimize them. Memory accesses are measured through instruction-based sampling (IBS) that relies on hardware support and requires using a kernel module, constraining the tool to the Linux/AMD platforms. DINAMITE can be extended with a native plugin or Spark kernel to generate the same information. It sacrifices performance over accuracy and portability as it does not rely on hardware support for instruction sampling.
 
Similar to Memprof, DProf  \cite{pesterev2010locating} uses IBS to acquire memory traces to locate data types that stress the cache. Programmers can view data type statistics and how they behave in the cache, what data types generate the most cache accesses and misses, and the most common functions that operate on these data types. DProf required changes to the operating system, which is significant barrier for its adoption. The flexibility of DINAMITE enabled us to plug in a cache simulator into the framework and to generate the same information as DProf, but with greater flexibility to add new analysis and without attachment to a particular operating system or hardware. 

Other tools provide more source-level detail but are slower and less flexible. Memspy \cite{martonosi1992memspy} provides rich information on program execution, including the execution time, miss rate, and memory stall time broken down by code and data. Details are tracked by executing the application through a memory simluator and instrumention through a preprocessor, which is not  as portable as LLVM. Memspy reported a 22$\times$-58$\times$ slowdown in execution time.  DINAMITE's slowdowns are competitive with modern instrumentation frameworks and provide a pluggable framework for all kinds of data analyses.

Like MACPO \cite{rane2012enhancing}, our tool instruments data accesses at the compiler level instead of the binary level to keep source-level information. To combat overhead, MACPO limits instrumentation to "snapshots" of program execution, that are staggered in an attempt to capture complete program behaviour. Trace size is reduced by limiting instrumentation not only to "snapshots" but also to non-scalar data types. Similarily Dprof and Memprof use IBS and only output the most commonly accessed data and their cache statistics. DINAMITE instruments all memory accesses inviting unlimited flexibility of analysis at the expense of higher runtime overhead.

%% file: futurework.tex
\section{Future work and conclusions}\label{sec:future_work}

We described the implementation of DINAMITE and discussed the performance implications of our design choices: a fine-grained compiler-based instrumentation and a flexible analysis framework. Our detailed breakdown of the costs involved in doing instrumentation of memory accesses leads us to conclude that this kind of design is not only feasible, but allows for a great level of detail in the generated traces, while keeping the slowdown comparable to the state-of-the-art instrumentation frameworks. We introduced a novel fusion of instrumentation and stream processing that eliminates the need for storing traces and provides an easy to use Spark Streaming API for analysis purposes. Finally we demonstrated the utility of DINAMITE by performing three \svetozar{different} types of analysis that were difficult, impossible, or constrained to a certain OS/hardware platform with the previously available tools.

In future work, we plan to expand on the  kinds of analysis that can be done on access traces using a streaming framework. Spark Streaming currently buffers incoming log records and processes them at the expiration of a configurable timeslice. In our experience, the timeslice must be rather large (e.g., one second) to avoid performance problems with Spark, but with such a large timeslice the batches contain hundreds of thousands of accesses. Adding support for a framework that is able to process a batch of records after it accumulates a specified \emph{number} of records would let us have finer granularity in our analysis. This kind of setup would allow discovering access patterns within small windows of time, which is important for certain optimizations.

Further, we plan to explore and optimize the slowdown of DINAMITE when using the full analysis pipeline with Spark Streaming. Finding the best way to integrate the log generation and analysis is an important factor in improving the overall productivity of engineers using our system.

Finally, our implementation of the cache simulator is rather simplistic. Adding a multi-level cache simulator such as Dinero IV~\cite{edler1998dinero} and adding more cache information to the logs would help improve understanding how different data organization and access patterns affect program efficiency.

%% file: machines.tex
\appendix[Hardware]

All hardware in our evaluation was performed on one of the following machines:
\begin{itemize}
    \item \emph{Machine A} - AMD Opteron 6272, 4 chips with 16 cores and 16MB of last level cache each, and 512GB of RAM
    \item \emph{Machine B} - AMD Opteron 2435, 2 chips with 6 cores 6MB of last level cache each, and 32GB of RAM 
\end{itemize}

%% file: main.bbl
\begin{thebibliography}{10}
\providecommand{\url}[1]{#1}
\csname url@samestyle\endcsname
\providecommand{\newblock}{\relax}
\providecommand{\bibinfo}[2]{#2}
\providecommand{\BIBentrySTDinterwordspacing}{\spaceskip=0pt\relax}
\providecommand{\BIBentryALTinterwordstretchfactor}{4}
\providecommand{\BIBentryALTinterwordspacing}{\spaceskip=\fontdimen2\font plus
\BIBentryALTinterwordstretchfactor\fontdimen3\font minus
  \fontdimen4\font\relax}
\providecommand{\BIBforeignlanguage}[2]{{%
\expandafter\ifx\csname l@#1\endcsname\relax
\typeout{** WARNING: IEEEtran.bst: No hyphenation pattern has been}%
\typeout{** loaded for the language `#1'. Using the pattern for}%
\typeout{** the default language instead.}%
\else
\language=\csname l@#1\endcsname
\fi
#2}}
\providecommand{\BIBdecl}{\relax}
\BIBdecl

\bibitem{ailamaki1999dbmss}
A.~Ailamaki, D.~J. DeWitt, M.~D. Hill, and D.~A. Wood, ``Dbmss on a modern
  processor: Where does time go?'' in \emph{VLDB" 99, Proceedings of 25th
  International Conference on Very Large Data Bases, September 7-10, 1999,
  Edinburgh, Scotland, UK}, no. DIAS-CONF-1999-001.

\bibitem{ferdman2012clearing}
M.~Ferdman, A.~Adileh, O.~Kocberber, S.~Volos, M.~Alisafaee, D.~Jevdjic,
  C.~Kaynak, A.~D. Popescu, A.~Ailamaki, and B.~Falsafi, ``Clearing the clouds:
  a study of emerging scale-out workloads on modern hardware,'' in \emph{ACM
  SIGPLAN Notices}, vol.~47, no.~4.

\bibitem{lattner2005automatic}
C.~Lattner and V.~Adve, ``Automatic pool allocation: improving performance by
  controlling data structure layout in the heap,'' in \emph{ACM SIGPLAN
  Notices}, vol.~40, no.~6.

\bibitem{lachaize2012memprof}
R.~Lachaize, B.~Lepers, and V.~Qu{\'e}ma, ``Memprof: A memory profiler for numa
  multicore systems,'' in \emph{Presented as part of the 2012 USENIX Annual
  Technical Conference (USENIX ATC 12)}.

\bibitem{pesterev2010locating}
A.~Pesterev, N.~Zeldovich, and R.~T. Morris, ``Locating cache performance
  bottlenecks using data profiling,'' in \emph{Proceedings of the 5th European
  conference on Computer systems}.

\bibitem{boyd2010analysis}
S.~Boyd-Wickizer, A.~T. Clements, Y.~Mao, A.~Pesterev, M.~F. Kaashoek,
  R.~Morris, N.~Zeldovich \emph{et~al.}, ``An analysis of linux scalability to
  many cores.'' in \emph{OSDI}, vol.~10, no.~13.

\bibitem{goto2008anatomy}
K.~Goto and R.~A. Geijn, ``Anatomy of high-performance matrix multiplication,''
  \emph{ACM Transactions on Mathematical Software (TOMS)}, vol.~34, no.~3.

\bibitem{yoon2005cache}
S.-E. Yoon, P.~Lindstrom, V.~Pascucci, and D.~Manocha, ``Cache-oblivious mesh
  layouts,'' in \emph{ACM Transactions on Graphics (TOG)}, vol.~24, no.~3.

\bibitem{bruening2003infrastructure}
D.~Bruening, T.~Garnett, and S.~Amarasinghe, ``An infrastructure for adaptive
  dynamic optimization,'' in \emph{Code Generation and Optimization, 2003. CGO
  2003. International Symposium on}.

\bibitem{de2010new}
A.~C. de~Melo, ``The new linux’perf’tools,'' 2010.

\bibitem{bitzes2014overhead}
G.~Bitzes and A.~Nowak, ``The overhead of profiling using pmu hardware
  counters,'' \emph{CERN openlab report}, 2014.

\bibitem{itzkowitz2003memory}
M.~Itzkowitz, B.~J. Wylie, C.~Aoki, and N.~Kosche, ``Memory profiling using
  hardware counters,'' in \emph{Supercomputing, 2003 ACM/IEEE Conference}.

\bibitem{nethercote2007valgrind}
N.~Nethercote and J.~Seward, ``Valgrind: a framework for heavyweight dynamic
  binary instrumentation,'' in \emph{ACM Sigplan notices}, vol.~42, no.~6.

\bibitem{luk2005pin}
C.-K. Luk, R.~Cohn, R.~Muth, H.~Patil, A.~Klauser, G.~Lowney, S.~Wallace, V.~J.
  Reddi, and K.~Hazelwood, ``Pin: building customized program analysis tools
  with dynamic instrumentation,'' in \emph{Acm sigplan notices}, vol.~40,
  no.~6.

\bibitem{lattner2004llvm}
C.~Lattner and V.~Adve, ``Llvm: A compilation framework for lifelong program
  analysis \& transformation,'' in \emph{Code Generation and Optimization,
  2004. CGO 2004. International Symposium on}.

\bibitem{zaharia2012discretized}
M.~Zaharia, T.~Das, H.~Li, S.~Shenker, and I.~Stoica, ``Discretized streams: an
  efficient and fault-tolerant model for stream processing on large clusters,''
  in \emph{Presented as part of the}, 2012.

\bibitem{chilimbi1999cache}
T.~M. Chilimbi, B.~Davidson, and J.~R. Larus, ``Cache-conscious structure
  definition,'' in \emph{ACM SIGPLAN Notices}, vol.~34, no.~5.

\bibitem{wt}
\BIBentryALTinterwordspacing
(2016) Wired tiger: making big data roar. [Online]. Available:
  \url{http://www.wiredtiger.com/}
\BIBentrySTDinterwordspacing

\bibitem{wt-mongo-db}
\BIBentryALTinterwordspacing
(2016) Wiredtiger storage engine. [Online]. Available:
  \url{https://docs.mongodb.com/manual/core/wiredtiger/}
\BIBentrySTDinterwordspacing

\bibitem{wt-pull-request}
\BIBentryALTinterwordspacing
(2016) Wt-2029 improve scalability of statistics. [Online]. Available:
  \url{https://github.com/wiredtiger/wiredtiger/pull/2102}
\BIBentrySTDinterwordspacing

\bibitem{drepper2007every}
U.~Drepper, ``What every programmer should know about memory,'' \emph{Red Hat,
  Inc}, vol.~11.

\bibitem{1_henning_2006}
J.~L. Henning, ``Spec cpu2006 benchmark descriptions,'' \emph{ACM SIGARCH
  Computer Architecture News}, vol.~34, no.~4, 2006.

\bibitem{bienia11benchmarking}
C.~Bienia, ``Benchmarking modern multiprocessors,'' Ph.D. dissertation,
  Princeton University, January 2011.

\bibitem{lee1997bit}
H.~B. Lee and B.~G. Zorn, ``Bit: A tool for instrumenting java bytecodes.'' in
  \emph{USENIX Symposium on Internet technologies and Systems}.

\bibitem{dice2013scalable}
D.~Dice, Y.~Lev, and M.~Moir, ``Scalable statistics counters,'' in
  \emph{Proceedings of the twenty-fifth annual ACM symposium on Parallelism in
  algorithms and architectures}.

\bibitem{malladi2009using}
R.~K. Malladi, ``Using intel{\textregistered} vtune™ performance analyzer
  events/ratios \& optimizing applications,'' \emph{http:/software. intel.
  com}, 2009.

\bibitem{zhao2011dynamic}
Q.~Zhao, D.~Koh, S.~Raza, D.~Bruening, W.-F. Wong, and S.~Amarasinghe,
  ``Dynamic cache contention detection in multi-threaded applications,'' in
  \emph{ACM SIGPLAN Notices}, vol.~46, no.~7.

\bibitem{zhao2010umbra}
Q.~Zhao, D.~Bruening, and S.~Amarasinghe, ``Umbra: Efficient and scalable
  memory shadowing,'' in \emph{Proceedings of the 8th annual IEEE/ACM
  international symposium on Code generation and optimization}.

\bibitem{liu2011sheriff}
T.~Liu and E.~D. Berger, ``Sheriff: precise detection and automatic mitigation
  of false sharing,'' \emph{ACM SIGPLAN Notices}, vol.~46, no.~10.

\bibitem{martonosi1992memspy}
M.~Martonosi, A.~Gupta, and T.~Anderson, ``Memspy: Analyzing memory system
  bottlenecks in programs,'' in \emph{ACM SIGMETRICS Performance Evaluation
  Review}, vol.~20, no.~1.

\bibitem{rane2012enhancing}
A.~Rane and J.~Browne, ``Enhancing performance optimization of multicore chips
  and multichip nodes with data structure metrics,'' in \emph{Proceedings of
  the 21st international conference on Parallel architectures and compilation
  techniques}.

\bibitem{edler1998dinero}
J.~Edler and M.~D. Hill, ``Dinero iv trace-driven uniprocessor cache
  simulator,'' 1998.

\end{thebibliography}
